\documentclass[titlepage,11pt]{article}

\usepackage{graphicx}

\usepackage[myheadings]{fullpage}
\usepackage{pmetrika}
\usepackage{pmbib}
\usepackage[natbibapa]{apacite}
\usepackage{amsmath,amsfonts,amssymb,bbm}
\usepackage{submit}
\usepackage{multirow}
\usepackage{float}

\usepackage{booktabs}
\usepackage{siunitx}
\usepackage{makecell}

\begin{document}

\begin{titlepage}

\title{Applying the Network Item Response Model to Student Assessment Data}


\author{Alex Brodersen}
\affil{University of Notre Dame}

\author{Ick Hoon Jin}
\affil{Yonsei University}

\author{Ying Cheng}
\affil{University of Notre Dame}

\author{Minjeong Jeon}
\affil{University of California - Los Angeles}


\vspace{\fill}\centerline{\today}\vspace{\fill}

\thanks{The authors would like to thank the anonymous referees, an Associate
Editor and the Editor for their constructive comments that improved the
quality of this paper.}

\contact{Correspondence concerning this article should be addressed to Alex Brodersen, 
Department of Psychology, University of Notre Dame, Notre Dame, IN. 46556. 
E-Mail: abroders@nd.edu}

\comment{Ick Hoon Jin was partially supported by the Yonsei University Research Fund of 2019-22-0210 and by Basic Science Research Program through the National Research Foundation of Korea (NRF 2020R1A2C1A01009881). Ying Cheng was supported by the National Science Foundation CAREER award (Grant \#DRL-1350787)}

\end{titlepage}

\setcounter{page}{2}
\vspace*{2\baselineskip}

\RepeatTitle{Applying the Network Item Response Model to Student Assessment Data}\vskip3pt

\linespacing{1.5}
\abstracthead
\begin{abstract}
This study discusses an alternative tool for modeling student assessment data. The model constructs networks from a matrix item responses and attempts to represent these data in low dimensional Euclidean space. This procedure has advantages over common methods used for modeling student assessment data such as Item Response Theory because it relaxes the highly restrictive local-independence assumption. This article provides a deep discussion of the model and the steps one must take to estimate it. To enable extending a present model by adding data, two methods for estimating the positions of new individuals in the network are discussed. Then, a real data analysis is then provided as a case study on using the model and how to interpret the results. Finally, the model is compared and contrasted to other popular models in psychological and educational measurement: Item response theory (IRT) and network psychometric Ising model for binary data. 
\end{abstract}\vspace{\fill}\pagebreak


\section{Introduction}

Item response theory models commonly posit the ``local'' independence assumption that patterns of response vectors to items administered in achievement tests are due solely to probabilistic functions of some latent ability vector. This assumption is quite strong in that model modifications must be made to accommodate group clusters, such as students clustered within classrooms. It also necessitates careful evaluation such that item parameters are invariant between student attributes such as age, gender, race, and primary language. A method for analyzing complex item response data utilizing an item response network was recently proposed by \citet{jin2019doubly}. The approcate, termed the network item response model (NIRM) is an exploratory method that creates a multiplex of item and person networks and estimates positions of both items and persons in a single latent space. The model was developed such that it relaxes the common local independence assumption in traditional item response models. Thus, clustering of item responses due to extraneous factors do not have negative implications for estimation, but rather are captured and represented by relative distances in an item latent space.

\subsection{Description of Network Item Response Models}

Latent space modeling \citep{hoff2002latent, handcock2007model, raftery2012fast} approaches have been used successfully in the analysis of social networks to represent the networks as a set of latent positions in Euclidean space. Recently, \citet{gollini2016joint} extended the latent space modeling literature by utilizing multiple network views from the same latent space to be used to estimate the latent space. The network item response model adopts this approach to represent persons and items in the same latent space. There are two specifications that enable this approach. First, the network item response model conceptualizes pairwise person and item data as a set of networks. Typical item response datasets have the form $\mathbf{X}_{n \times p}$ where $x_{ki}$ represents the binary response of the $k$th person to the $i$th item. The network item response model utilizes a different unit of analysis than the typical rows and columns representation of the data. Instead, fitting the model requires constructing two sets of binary adjacency matrices. The two sets contain $n$ $p \times p$ and $p$ $n \times n$ binary adjacency matrices, herafter refered as networks. There is one network for each person and one network for each item. The elements of these networks consist of a representation of the interactions of persons within items and items within persons. There are multiple choices for the construction of these networks, which is discussed in a later section.

The full data likelihood for the network item response model is the joint probability of observing both sets of networks $Y_{i,pxp}$ and $U_{k,pxp}$, which is a function of the latent space $\mathbf{Z}^{d}$ (where $d$ is the number of dimensions), intercept terms for each item $\mathbf{\beta}_{1xp}$, and intercept terms, $\theta$, for each person.

$$
\begin{aligned}
  P(\mathbf{Y}, \mathbf{U} | \mathbf{Z}, \boldsymbol{\beta}, \boldsymbol{\theta})&=\prod_{i=1}^{p} P\left(\mathbf{Y}_{i} | \mathbf{Z}, \beta_{i}\right) \prod_{k=1}^{n} P\left(\mathbf{U}_{k} | \mathbf{Z}, \theta_{k}\right) \\
  &=\prod_{i=1}^{p} \prod_{k \neq l} \frac{\exp \left(\beta_{i}-\left\|\mathbf{z}_{k}-\mathbf{z}_{l}\right\|\right)^{y_{i, k t}}}{1+\exp \left(\beta_{i}-\left\|\mathbf{z}_{k}-\mathbf{z}_{l}\right\|\right)} \prod_{k=1}^{n} \prod_{i \neq j} \frac{\exp \left(\theta_{k}-\left\|\mathbf{w}_{i}-\mathbf{w}_{j}\right\| \right)^{u_{k, i j}}}{1+\exp \left(\theta_{k}-\left\|\mathbf{w}_{i}-\mathbf{w}_{j}\right\|\right)}&
\end{aligned}
$$

The second specification is that the network item response model enables a joint modeling of latent item and person spaces. To accomplish this requires a mapping from one space to another. \citet{jin2019doubly} used a mapping that assumed the item space is a function of the person space. Specifically, the item latent positions were equal to the average person positions of those who endorsed the item (e.g. answered it correctly for educational tests, or responded in the affirmative for psychological tests). Other mappings are possible, and this will also be discussed further in a later section.

$$
\begin{aligned}
  P(\mathbf{Y}, \mathbf{U} | \mathbf{Z}, \boldsymbol{\beta}, \boldsymbol{\theta})&=\prod_{i=1}^{p} P\left(\mathbf{Y}_{i} | \mathbf{Z}, \beta_{i}\right) \prod_{k=1}^{n} P\left(\mathbf{U}_{k} | \mathbf{Z}, \theta_{k}\right) \\
  &=\prod_{i=1}^{p} \prod_{k \neq l} \frac{\exp \left(\beta_{i}-\left\|\mathbf{z}_{k}-\mathbf{z}_{l}\right\|\right)^{y_{i, k t}}}{1+\exp \left(\beta_{i}-\left\|\mathbf{z}_{k}-\mathbf{z}_{l}\right\|\right)} \prod_{k=1}^{n} \prod_{i \neq j} \frac{\exp \left(\theta_{k}-\left\|f_{i}(\mathbf{z})-f_{j}(\mathbf{z})\right\| \right)^{u_{k, i j}}}{1+\exp \left(\theta_{k}-\left\|f_{i}(\mathbf{z})-f_{j}(\mathbf{z})\right\|\right)}&
\end{aligned}
$$

A critical point in differentiating NIRM from other models is the unit of analysis, which is a pairwise summary of item and person information - not individual person level data itself. In general, it is not possible to completely recover the original data matrix from it's network representation, and therefore it is not directly equivalent to many other model-based procedures for investigating psychological and educational data, such as Item Response Theory (IRT) or network psychometric models such as the Ising Model for binary data or Gaussian Graphical Model (GGM) for continuous data.

\subsection{Objectives}

To date there only two articles regarding the NIRM approach to blending networks and item response models. The first was the introduction of the model \citep{jin2019doubly} and the second extended the model the model to allow for hierarchical dependencies \citep{jin2018hierarchical}. The depth of technical details in those articles prevented a thorough case study in the use of this model. This article has four aims. First, this article will provide a deep description of all analysis decisions that need to be made prior to using NIRM, including some new decision points that had not been considered in \citet{jin2019doubly}. Second, we will provide solutions for incorporating new data into and existing NIRM. Adding new individuals, new items, or both are common in educational assessment and this article will enable these activities for a previously estimated NIRM. For example, in IRT these processes are termed ability estimation and linking. However, the analogues to ability estimation and equating have not been previously been discussed in NIRMs. Third, we will then present analysis of a dataset using NIRM. The case study shows the utility of the NIRM by demonstrating its use on an example data set containing item-level responses to a high school statistics students assessment, which was administered to classrooms across several different schools. This analysis will include providing interpretation from known properties of the items (e.g. content area) as well as illuminate potential issues with the item responses. We provide the previous two sections separately because the real data analysis represents one set of a combination of analysis decisions and a discussion of each decision intertwined with the analysis would confound the aims of both. Finally, this article will describe the distinction between other commonly used models for item responses in education and psychology. Specifically, this article will compare and contrast between the NIRM, the Ising Model, IRT Models.

\section{Practical Recommendations when Using NIRM} \label{usingnirm}

Prior to using NIRM, there are several determinations that need to be made. The first of these is the selection of the dimensionality of the latent space. Then, one must decide how to construct the item networks, which is effectively a choice on defining interesting margins in a $2 \times 2$ contingency table for every pair of persons and items. The final step is selecting a linkage between the item latent space and the person latent space. The remainder of this section will discuss methods for estimation of NIRM models, optional and necessary steps for post-processing the results, and a guide for interpreting and visualizing the model.

\subsection{Selection of the Dimension of the Latent Space}
This step is a conceptually similar problem to selecting the number of factors in factor analysis or IRT. Similar to the number of factors problem, there is no satisfactory solution because the true data generation process is likely much more complex than a simple function of a low-dimension latent space. However, commonly applied rules for selecting the number of latent dimensions in factor analysis such as the scree test \citep{cattell1966scree}, parallel analysis \citep{horn1965rationale}, and the ``eigenvalue great than one'' rule \citep{kaiser1960application} do not apply to NIRM. Likewise, NIRM does not presently have analogues to common measures of fit in factor analysis or structural equation modeling such as TLI \citep{tucker1973reliability},  standardized root-mean-square residual (SRMR), or RMSEA \citep{steiger1980paper,browne1993alternative}. That said, there are two considerations for selecting the number of dimensions in the latent Euclidean space. The first criterion is ease of interpretation of the resultant latent space. One purpose of NIRM is to project summary information about pairwise responses of persons and items in order to obtain information about item properties. To this end, it is clear that a preference for simplification and selecting the minimum number of dimensions is a worthy goal. Therefore it is often reasonable to select 2 or 3 dimensions for the purpose of easing the process visualizing the latent spaces. The second criterion is the risk of ``underfactoring''. NIRM is conceptually similar to multi-dimensional scaling in that it attempts to identify an optimal representation of \textit{distances} between persons, rather than quantify magnitudes of a latent trait as in IRT. If the selected number of dimensions is too low for this purpose, the visualized distances may not be an accurate representation of the true distances. The implications of this have been acknowledged in the literature many times over, but it bears repeating here. While approaches from the factor analytic literature may not provide useful criterion, we acknowledge that due to conceptual similarities, approaches from parameterized versions of multidimensional scaling give insight into selecting the number of dimensions for NIRM.

Specifically, we note it is possible to place a diffuse prior over the integer number of dimensions up to a pre-specified maximum and select the choice that optimizes a corresponding information criterion \citep{oh2001bayesian}. However, the application of this is not straightforward. Relatedly, \citet{oh2001bayesian} note several issues with bayesian MDS that are directly applicable to NIRM. Namely, for identical Euclidean distances, the coordinates of $\mathbf{Z}$ must get closer to the origin as $d$ increases, unless all the extra coordinates are equal to 0. This is a direct consequence of the curse of dimensionality \citep{zimek2012survey}. Overall, this is a complex issue, but starting with a low number of dimensions such as 2 or 3 can serve as a good rule of thumb.

\subsection{Construction of Item Networks}

As previously mentioned, the unit of analysis for NIRM is different from IRT or Ising Models. From the original response dataset, we construct two sets of binary networks for pairwise representations of persons and items. That is, for each item $i$, we have an $n \times n$ binary network, $\mathbf{Y}_{i, n \times n}$. Likewise, for each person $k$, we have a $p \times p$ network, $\mathbf{U}_{k, p \times p}$. Each element of networks $\mathbf{Y}$ or $\mathbf{U}$ is constructed as a function of the appropriate elements of the original data matrix, $\mathbf{X}_{n \times p}$. For example, from below, in network $Y_i$, the element in the $k^{th}$ row and the $l^{th}$ column, (i.e. element $y_{i, k, l}$) is a function of elements from the $k^{th}$ and $l^{th}$ rows of the $i^{th}$ column in data matrix $\mathbf{X}$ (i.e. $x_{k i} \text{ and } x_{l i}$).

\begin{equation}
  \mathbf{Y}_{i, n \times n}=\left\{y_{i, k l}\right\}=f(x_{k i}, x_{l i}) \quad \text { and } \quad \mathbf{U}_{k, p \times p}=\left\{u_{k, i j}\right\}=f(x_{k i}, x_{k j})
\end{equation}

The original article on NIRM discussed only a single method for constructing the networks from a matrix of item responses. It conceptualized each element of the matrix as an encoding of whether or not both elements of comparison took a value of 1. That is, each element of one of the networks was the product of two elements of the original data matrix, represented in Equation \ref{eq:prod}.

\begin{equation}
  \label{eq:prod}
\mathbf{Y}_{i, n \times n}=\left\{y_{i, k t}\right\}=\left\{x_{k i} x_{l i}\right\} \quad \text { and } \quad \mathbf{U}_{k, p \times p}=\left\{u_{k, i j}\right\}=\left\{x_{k i} x_{k j}\right\}
\end{equation}
  
That is, because of the binary nature of the data, this product takes a value of 1 when both elements are 1, and 0 otherwise. While described as an interaction between the two variables, taking the product is simply one method of classifying pairwise responses into a set of operational taxonomic units \citep{sokal1961principles}.

\tablehere{\ref{tab:otu}}
Strictly speaking, any substantively interesting enumeration of the elements of Table  \ref{tab:otu} can be used as an element of analysis. In the original article, the operational unit was taken to be positively matched concordant pairs. This lead to the above specification. The psychological interpretation of this enumeration might be that an affirmative response indicates presence of the trait, but the absence of an affirmative response doesn't necessarily indicate absence of the trait. This enumeration is in fact quite common in psychological classifications as diagnostic criteria may declare a threshold on the presence of a fixed number of criteria to imply a diagnosis but the absence of any of the non-endorsed criteria are not counter-evidence to a diagnosis. Another example of this is the literature on positive and negative affect in which the absence of one does not imply the presence of the other. This selection of how to enumerate the data is somewhat analogous to the selection of ${0,1}$ encoding or ${-1,1}$ encoding in Ising models \citep{haslbeck2018interpreting} with the distinction being the selection within the Ising model results in statistically equivalent models but in NIRM it does not.

In other fields, such as educational measurement, the presence or absence of a correct response is taken as evidence a high or low magnitude of a trait, respectively. Here, we are coding similarity of responses as an indicator function that takes value 1 if respondents have the same endorsement and 0 otherwise. For binary data, this may be mathematically represented as below.

\begin{equation}
  \label{eq:pow}
\mathbf{Y}_{i, n \times n}=\left\{y_{i, k t}\right\}=\left\{x_{k i}^{x_{l i}} x_{l i}^{x_{k i}} \right\} \quad \text { and } \quad \mathbf{U}_{k, p \times p}=\left\{u_{k, i j}\right\}=\left\{x_{k i}^{x_{k j}} x_{k j}^{x_{k i}}\right\}
\end{equation}

Furthermore, while the network of item responses must contain binary information, there is nothing that prevents dummy-coding a single item or collection of items such that meaningful binary encodings are including in the item response networks. One example of this would be ``multiple-mark'' or ``multiple TRUE/FALSE'' item type \citep{frisbie1982relative}, which contains a clustered set of check-box type responses with a ``mark-all-that-apply'' type prompt. Traditionally, item response analysis techniques have struggled with this item type for three reasons. First, the clustered set of options nested under a single prompt is a prototypical example of the type of items that cause local dependence \citep{yen1984effects}. Second, the common choice for analyzing these data types is to either treat the sum score of the individually marked elements as a single ordinal item or to apply a scoring rule that accounts for or penalizes over/under marking and then again treat the outcome as a single ordinal item. Clearly, this ignores information encoded in the pattern of responses. The third struggle with the item type follows from the first and second. If analyzing the sub-items separately and analyzing a sum score are both sub-optimal, then the next best correct solution would be to analyzing the whole patterns of responses from the mark-M item. However, the number of patterns for this item type is $2^k$ where $k$ is the number of sub-items. For most seemingly reasonable number of sub-items, $2^k$ is much too large to expect a non-zero number of observations of each pattern.

\subsection{Choosing the linkage between latent spaces} \label{linkage}
The item and person latent spaces are jointly estimated, but there is a misalignment in the dimensionality of the item networks and person networks. \citet{jin2019doubly} resolve this problem by assuming that the one space can be defined as a function of the other. Logically, this makes sense as both networks are constructed from the same dataset, $\mathbf{X}$.  There are many possible choices of linkages between the two latent spaces. Yet, \citet{jin2019doubly} mentioned only a single method of linking the two, which defined the latent position of item $i$, $\mathbf{w}_i$, to be the average of the latent position of the respondents who answered item $i$ correctly. That is, the chosen the mathematical link is:

$$\mathbf{w}_{i}=f_{i}(\mathbf{Z})=\sum_{k=1}^{n} \frac{x_{k i} \mathbf{z}_{k}}{\sum_{k=1}^{n} x_{k i}}$$

To give further intuition on this choice of link, consider that each person $k \in 1,\ldots, n$ contributes their own item network to the analysis. This network contains a binary similarity measure for each item pair within individuals and both the persons tendency have elements of their item network take values of 1 (i.e., $\theta$) and the distance between items determine the probability of a link. Then, each person's network contributes to determining the distance between the items, controlling for a persons tendency to answer pairs of items similarly. Thus, it is reasonable that an average of the persons who answered an item correctly defines the location of that item.

However, there is nothing preventing additional methods of resolving the dimensional mismatch and linking the two spaces. A clear alternative would be the assumption that this influence goes the other direction. That is, that the respondent latent space is a function of the item latent space and a reasonable approximation of a persons latent position is the average of the positions of the items which they answers correctly.

$$\mathbf{z}_{i}=f_{i}(\mathbf{W})=\sum_{k=1}^{p} \frac{x_{k i} \mathbf{w}_{k}}{\sum_{k=1}^{p} x_{k i}}$$

This formulation may not always be appropriate. When the number of items is small (e.g., $p < 5$) and many items have low proportions of positive endorsement, the accuracy of estimating the positions of items is severely limited because few individuals are contributing to estimating the positions of the items.

\subsection{Estimation}

\citet[][pp. 244]{jin2019doubly} provides a detailed discussion of Bayesian estimation of NIRM vis a Metropolis-Hastings algorithm. Unfortunately, this algorithm is either $\mathcal{O}(n^2p)$ or $\mathcal{O}(p^2n)$ depending on whether the person space is a function of the item space or the item space is a function of the person space, respectively. Preliminary investigations into variational approximations to this estimation procedure via an EM algorithm have found that this approach may be too slow to be useful. This might appear contrary to the results of \citet{gollini2016joint}, but the distinction between that context and the NIRM is that the applied analysis in \citet{gollini2016joint} jointly modeled three network views while NIRM requires modeling at least $n + p \gg 3$ network views. Knowing this, the EM algorithm becomes $\mathcal{O}(((n * q) ^ d)^2p)$ where $q$ is the selected number of quadrature points and $d$ is the selected dimension of the NIRM. This grows very quickly and becomes unfeasible even in moderately sized samples. Thus, the Bayesian procedure remains the method of choice for estimating NIRM.

In the Bayesian procedure, one must select prior information for $\theta_k$, $\beta_i$, $\mathbf{z}_k$ or $\mathbf{w}_i$. Normal priors are suggested for all first order parameters, i.e.

$$p(\theta_k) \sim N(0,\sigma^2_{\theta})$$
$$p(\beta_i) \sim N(0,\sigma^2_{\beta})$$

and 

$$p(\mathbf{z}_k|\sigma_z^2) \sim N(\mathbf{0},\sigma^2_{z}\mathbf{I}_d)$$
$$p(\mathbf{w}_i|\sigma_w^2) \sim N(\mathbf{0},\sigma^2_{w}\mathbf{I}_d)$$

Note that the choice of $\mathbf{0}$ as a prior mean for $\mathbf{z}_k$ or $\mathbf{w}_i$ is arbitrary as the mean contributes no information to the distances between latent positions. Additionally, the choice of a diagonal matrix is arbitrary. However, we note that if this specification is made then $\sigma^2_{z}$ or $\sigma^2_{w}$ can be estimated by assigning a hyper prior. This provides useful information as the variance of the latent spaces can be viewed as a pseudo effect size of the model. That is, a large value of $\sigma^2_{z}$ results in larger distances between person and item positions in the network and proportionally lessens the influence of the marginal tendencies of 1's in the person networks ($\theta_k$) and item networks ($\beta_i$).

An Inv-gamma distribution is recommended to estimate the effects of the distances on the network. That is, we recommend 

$$p(\sigma^2_{z}) \sim \text{Inv-Gamma}(a_{\sigma^2_{z}},b_{\sigma^2_{z}})$$

and

$$p(\sigma^2_{w}) \sim \text{Inv-Gamma}(a_{\sigma^2_{w}},b_{\sigma^2_{w}})$$

At present, there has been little investigation into the effects of selection of hyper-parameters for this model. In other Bayesian analyses, it is common to select $a_{\sigma^2} = b_{\sigma^2}$ such that the prior expectation of $\sigma^2$ equals 1. While this does have influence over the contribution of the latent space, we suspect that said influence is minimal. This is because in a distance-based approach the amount of information about $\sigma^2$ grows with $n^2$ instead of $n$, even at low sample sizes this information can quickly overwhelm the prior.

\subsection{Post-processing}

Clearly in a Bayesian analysis it is important to check convergence of the parameter sampling chains. This is straightforward for the person and item intercept terms, ($\beta_i$ and $\theta_k$). However, because Euclidean distance is invariant under translation, rotation, and reflection, one must take extra care in diagnosing the results of the latent positions. First, analyzing the sampling chain of an individual's latent position without post-processing will likely show the strong serial correlations. This is because the lack of identification of the latent positions allow the latent space to freely transform during sampling. To correct for this, Procrustes matching can be applied to the MCMC samples. More appropriately, selecting pairs of individuals or items, calculating the distance between the pair at each sampling iteration, and plotting the resulting sampling chain of the distance can be used to diagnose sampling issues with the distance metrics. Naturally, one may opt to perform this procedure for a sample of pairs due to the large number of pairs, i.e. $\binom{n}{2}$.

For visualization purposes, it may be useful to rotate a matrix of latent positions. For example, the estimated $n \times d$ matrix of estimated latent person positions can be centered at the origin and rotated to its principal axes orientation \citep{borg2005modern}, giving the positions zero mean and a diagonal covariance matrix. Specifically, let $\hat{Z}$ be the matrix of estimated position. Then, let $\mathbf{Q}$ be the $d \times d$ matrix of eigenvectors of $\hat{Z}^\intercal\hat{Z}$. Preferably, the columns of $\mathbf{Q}$ are ordered by the magnitude of the corresponding eigenvalues.

Then, the transformation $\hat{Z}^{*} = \hat{Z}\mathbf{Q}$ rotates the latent space such that the dimensions are in decreasing order of their contribution to the observed networks. There are infinitely many other choices for rotation matrices, and similar topics have been evaluated in great depth in the exploratory factor analysis literature. However, many rotation methods have the aim of achieving a ``simple structure'' in which many factor loadings of a single item are zero. This is not a fruitful aim when rotating latent positions as a zero-valued position on a single dimension does not have any inherent meaning. However, target rotation \citep{browne1972oblique} may add insight into the latent positions for meaningfully specified target matrices. For example, specifying a target matrix such that certain subgroups of items obtain a meaningful pattern of positive and negative sign could create a rotation such that the orthant in which an item lies is substantively meaningful.

\subsection{Interpretation}

Interpretation for the person ($\theta_k$) and item ($\beta_i$) intercept terms partially depends on the network encoding selected in previous stages of the analysis. In any case, the intercept term reflects the density of the associated person or item networks, but of course that density has different interpretations depending on the encoding. If networks are constructed as in Equation \ref{eq:prod}, then $\theta_k$ reflects the tendency of person $k$ to answer pairs of items in the ``1'' category and $\beta_i$ reflects the tendency for pairs of individuals to both respond in the ``1'' category to item $i$. Conversely, if networks are constructed as in Equation \ref{eq:pow}, then $\theta_k$ reflects the tendency of person $k$ to answer pairs of items in a concordant manner (i.e. both ``0'' or both ``1'') and $\beta_i$ reflects the tendency for pairs of individuals to respond in a concordant manner to item $i$, again, both ``0'' or both ``1''. Typically, latent person positions $\mathbf{z}_k$ cannot be interpreted directly; they must be interpreted as a function of distance from either other persons or other items. In practice, one can inspect the distance from a person to a centroid of other persons or a centroid of items. In the context of a single classroom, an individual with a large distance to other students in the same classroom likely has largely different strengths and/or deficiencies on the material compared to other students. In psychology, large distances may be interpreted as individuals currently residing in a different developmental state \citep{jin2019doubly} than others, or perhaps having a largely different set of psychological symptoms than others. In the latter case, and given proper considerations for the source of the sample, clusters of individuals may represent relatively stable patterns of symptoms that may otherwise be in flux, i.e. a collection of low energy states.

\subsection{Visualization}

\begin{figure}[H]
\centerline{\includegraphics[scale=.3]{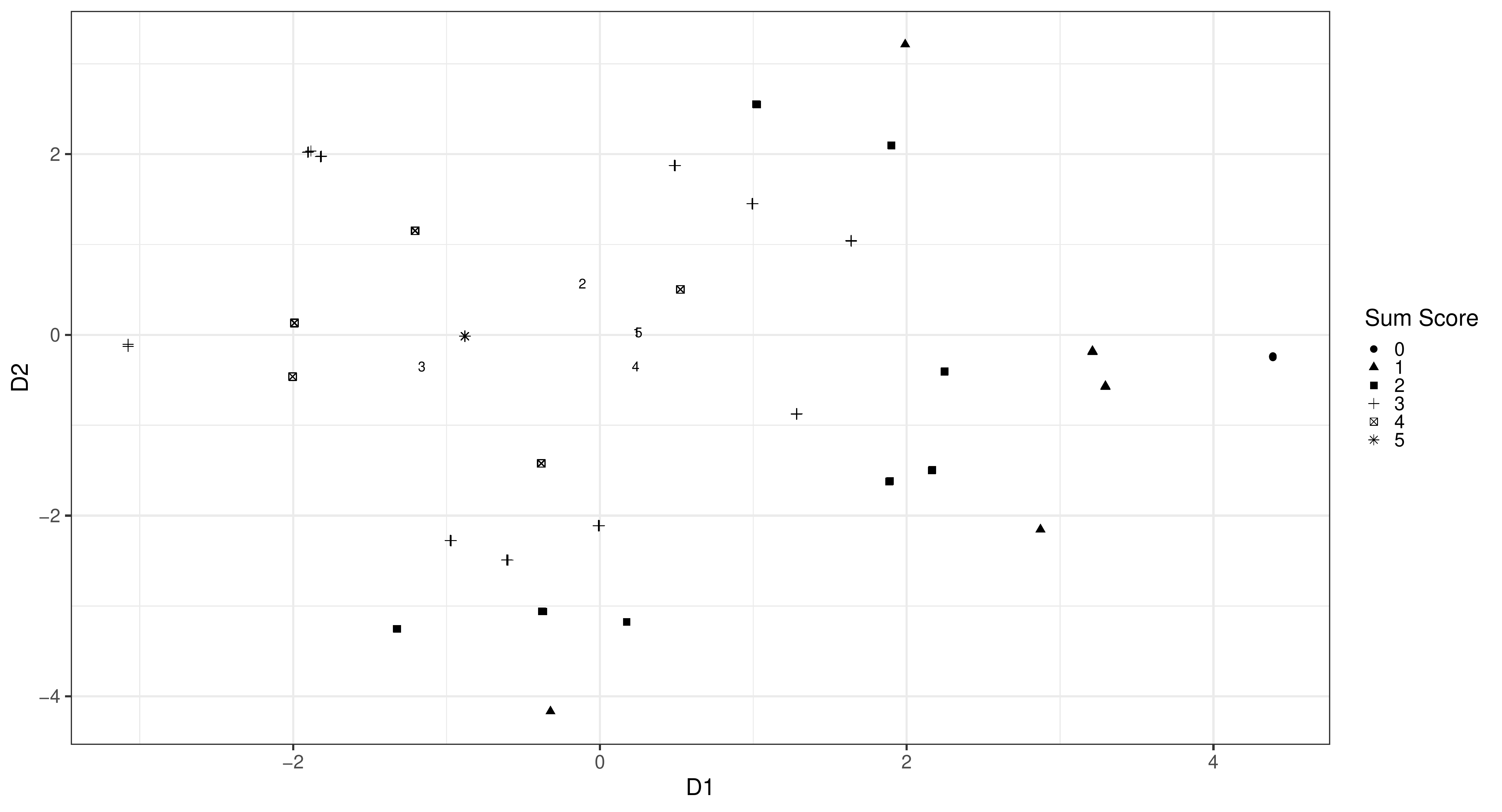}}
    \caption{Item and person positions for the LSAT 6 data. Note item 3 has the largest distance from the other items and that item 1 and item 5 have almost identical positions. Person positions are given shapes based on their some score; persons with with the same sum score can have different positions based on the individual pattern score.}
    \label{fig:lsatp}
\end{figure}

There are two ways to visualize the results of a NIRM. The first is to construct plots visualizing the two latent spaces, or perhaps a single plot superimposing the item latent space on the person latent space. And example of these can be seen in Figure \ref{fig:lsatp}, which represents the NIRM estimated from the well-known LSAT6 dataset. In these plots, one can attempt to interpret the positions of items and persons relative to each other. 

\figurehere{\ref{fig:lsatp}}

Importantly, large distances imply largely different patterns of responding. It is also important to visually inspect for clusters of items; these occur when portions of the sample have responded in a highly consistent manner to them. One the other side, ``outliers'' that are far away from others are also informative as these are items which display highly dis-regular responding relative to other items. This may be items that are irrelevant from a content perspective or may indicate a miskeyed item in educational testing.

\begin{figure}[H]
\centerline{\includegraphics[scale=.45]{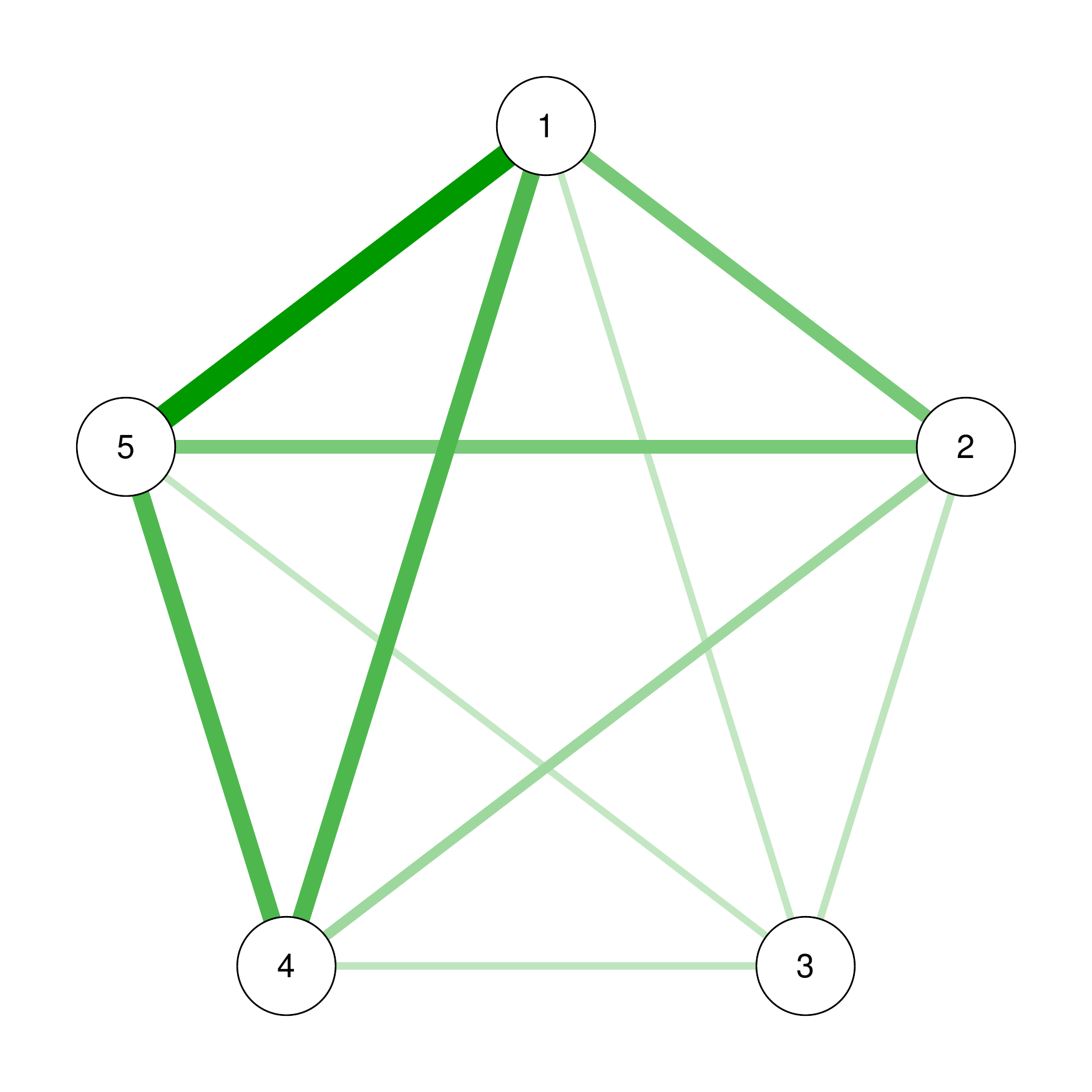}}
  \caption{Network for the LSAT 6 data with inverse exponentiated Euclidean distance similarity metrix ($s_1$).}
  \label{fig:lsatnetwork}
\end{figure}

Another approach to visualize a NIRM is to represent the items as a network of nodes where distances between items are the weights in the edges of the network. However, representing \textit{distances} as edges in the network may be misleading, and other network methods consider larger values (or at least, more positive values) to be indicative of higher ``connectedness''. Instead, it would be better to represent positions with a measure of similarity. An example of this of can be found in Figure \ref{fig:lsatnetwork}, and these plots can easily be generated with the \texttt{qgraph} package in R \citep{epskamp2012qgraph}.

\figurehere{\ref{fig:lsatnetwork}}

There are many choices for selecting an appropriate similarity metric. A natural first choice would be cosine similarity as it is related to Euclidean distance. However, unlike Euclidean distance cosine similarity is not invariant to translations and rotations of the latent space. Instead, one may select an order preserving distance to similarity transformation. For example,

$$s_1 = \frac{1}{e^{\left\|\mathbf{w}_i - \mathbf{w}_j \right\|}}$$
 is a suitable choice that maps Euclidean distance into the $\{0,1\}$ similarity space. Here, $w_i$ and $w_j$ are the latent positions of items $i$ and $j$. Another suitable choice for a similarity metric would be
 
 $$s_2 = 1 - \frac{\left\|\mathbf{w}_i - \mathbf{w}_j \right\|}{\text{max}_{i,j}(\left\|\mathbf{w}_i - \mathbf{w}_j \right\|)}$$

which also bounds the similarity metrix into $\{0,1\}$.

\section{Adding new data}

Given an estimated NIRM, it may be of interest to estimate the positions of new additions to the data. For example, one may want to estimate the positions of persons in a newly collected sample to compare positions to the old sample. This is analogous to ability estimation in IRT. Likewise, if data were collected on one or more new items from individuals in the same sample, it might be of interest to place those items in the existing latent space. This is analogous to linking item parameters of an IRT model.

\begin{figure}[H]
\centerline{\includegraphics[scale=.45]{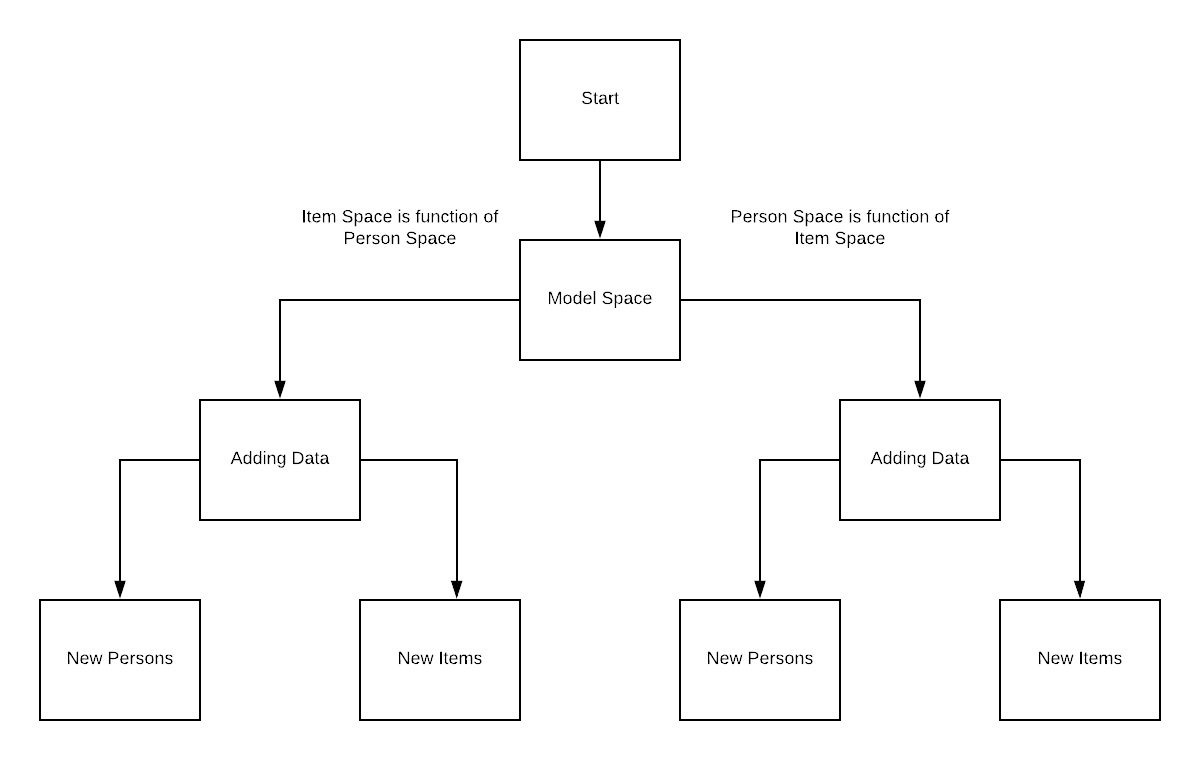}}
\caption{The 4 cases of estimating new latent positions in NIRM}
\label{fig:newcases}
\end{figure}

Unfortunately, estimating new positions in NIRM is not trivial. In first round estimation of NIRM, the latent space itself is estimated, which is a function of the sample and the items that were included in the analysis. The naive solution is to combine the old data and new data into a single dataset and estimate the full model on the whole data. This may not be a desirable solution as estimation is generally slow. Fortunately, there are some alternatives. These alternatives depend on the analysis choices that were made when estimating the original model. Specifically, the choice of the model space linkage and whether one is adding new items locations or new person locations. These are outlined in Figure \ref{fig:newcases}. 

\figurehere{\ref{fig:newcases}}

\subsection{Approximating new positions}

Importantly, there are some quickly calculated approximations to estimating new positions for person or items when the latent space has been estimated as a function of the opposite space. For example, if the person latent space has been estimated as a function of the item latent space, then there is a quick approximation for estimating the locations of a new sample of individuals. Similarly, if the item latent space has been estimated as a function of the person latent space, then there is also a quick approximation for estimating the locations of new items.

Relying on the link between the latent spaces, a new set of positions can be estimated by averaging the latent positions for the items which were answered correctly. This approximation is straightforward to calculate and when the number of items with correct responses is sufficiently large, it converges to the true value of the latent positions.

Therefore, if the item latent space is a function of the person latent space, this approximation may be used:

$$\hat{\mathbf{w}}_{i} = \sum_{k=1}^{n} \frac{x_{k i} \hat{\mathbf{z}}_{k}}{\sum_{k=1}^{n} x_{k i} }$$

Likewise, if the person latent space is a function of the items latent space, one can estimate new person positions with the following equation:

$$\hat{\mathbf{z}}_{k} = \sum_{i=1}^{p} \frac{x_{k i} \hat{\mathbf{w}}_{i} }{\epsilon + \sum_{i=1}^{p} x_{k i}}$$

Here, $\epsilon$ is an arbitrarily small number to prevent division by 0 which may occur if a response vector contains no correct responses. However, this approximation loses accuracy when very few items have been answered correctly and in the particular degenerate case of a zero sum-score will define the latent person vector to be $\mathbf{0}^d$. While seemingly not useful as an approximation in this case, zero correct responses indicates that there the participant is equally distant to all sets of items.

If these approximations are not useful the proceeding sections describe sampling procedures for estimating new positions assuming the latent spaces are fixed to their posterior estimates from the first sample. One benefit of these assumptions is the rotational and translational indeterminacy is no longer a factor and thus the samples do not require post-processing. 

In all cases, the selection of prior information can remain the same, with the exception of the hyper-prior on the variance of the latent spaces; this hyper-prior may be removed and the latent space variances can be fixed to their posterior estimates.

\subsection{Approximating new person intercepts}

Intercepts for new persons may not be substantively interesting as they mostly provide a correction for base rates (i.e. they are analogous to main effects) when estimating person positions. They have similar interpretations to the person and item location parameters of the Rasch model. Like the in the Rasch models, they provide little additional information above person sum scores, except to give an inverse logit transformed representation of the probability of concordant pairs. However, if these are of interest, a simple approximation is available. An approximatete intercepts for new persons are availible:

\begin{equation}
\label{eq:approxtheta}
  \tilde{\hat{\theta}}_{new} = \frac{\sum_{i} \mathbbm{1}{(\sum_j x_{ij} == \sum_j x_{new,j})}\hat{\theta}_i}{\sum_{i} \mathbbm{1}{(\sum_j x_{ij} == \sum_j x_{new,j})}}
\end{equation}

where $\mathbbm{1}{(\cdot)}$ is an indicator function that evaluates to 1 if the arguments are \texttt{TRUE} and 0 otherwise. This approximation is not useful when attempting to map a new sum score to $\theta$ when that value of the sum score was not observed in previous samples. Unfortunately, there is little to be done about these approximate estimators in that case. It is also not uncommon for estimators to give non-useful results for some special cases, such as patterns of all 1's or 0's in 2PL IRT models \citep[][pp. 26]{de2013theory}. However, we provide the approximations in hopes they may be useful in some circumstances, as it is expected these models may be used in an exploratory fashion. For cases when the approximations will not do, we describe a set of conditional posteriors that, with added assumptions, can be used to estimate parameters of persons and items for new data.

\subsection{Incorporating new persons}

When new data are added for new persons on the same items, we may wish to estimate the parameters corresponding to the new individuals. Here, we are attempting to estimate two new sets of information. The first is the new person intercept. Assuming that the new individuals are from the same latent space, the following conditional posterior for $\theta_{new}$ is provided below.

\begin{equation}
  \label{eq:condposttheta}
\pi\left(\theta_{new} | \mathbf{U}_{new}, \hat{\mathbf{Z}}\right) \propto \pi\left(\theta\right) \prod_{i \neq j} \frac{\exp \left(\theta_{new}-\left\|\hat{\mathbf{w}}_{i}-\hat{\mathbf{w}}_{j}\right\|\right)^{u_{k, i j}}}{1+\exp \left(\theta_{new}-\left\|\hat{\mathbf{w}}_{i}-\hat{\mathbf{w}}_{j}\right\|\right)}
\end{equation}

where $\hat{\mathbf{w}}_{i} = f_i(\hat{\mathbf{Z}})$ from the initial NIRM calibration. One also must estimation the locations for the new persons. The conditional posterior for new person positions $\mathbf{z}_{new}$ is given in equation \ref{eq:condpostz}. This requires both new and old datasets and parameter estimates from the previous models. Historically, it has been considered cumbersome to retain old data to estimate new portions of a model; however, with modern data retention practices we hope this is currently less of an issue. Unfortunately, it is also an unavoidable tradeoff with distance-based models.

\begin{equation}
  \label{eq:condpostz}
\begin{aligned}
  \pi\bigl(\mathbf{z}_{new}|&\mathbf{Y},\mathbf{U},\hat{\boldsymbol{\beta}},\boldsymbol{\theta},\hat{\boldsymbol{Z}}\bigr) \propto \pi\left(\mathbf{z}_{new} | \hat{\sigma}_{z}^{2}\right) \prod_{i=1}^{p} P\left(\mathbf{Y}_{i} | \mathbf{z}_{k}, \hat{\boldsymbol{Z}}, \beta_{i}\right) \prod_{k=1}^{n} P\left(\mathbf{U}_{k} | \hat{\mathbf{w}}_{i}, \hat{\theta}_{k}\right) \\
  & \propto \pi\left(\mathbf{z}_{new} | \hat{\sigma}_{z}^{2}\right)\prod_{i=1}^{p} \prod_{k \neq l} \frac{\exp \left(\beta_{i}-\left\|\mathbf{z}_{k}-\mathbf{z}_{l}\right\|\right)^{y_{i, k t}}}{1+\exp \left(\beta_{i}-\left\|\mathbf{z}_{k}-\mathbf{z}_{l}\right\|\right)} \prod_{k=1}^{n} \prod_{i \neq j} \frac{\exp \left(\hat{\theta}_{k}-\left\|\hat{\mathbf{w}}_{i}-\hat{\mathbf{w}}_{j}\right\| \right)^{u_{k, i j}}}{1+\exp \left(\hat{\theta}_{k}-\left\|\hat{\mathbf{w}}_{i}-\hat{\mathbf{w}}_{j}\right\|\right)}\\
\end{aligned}
\end{equation}

Similarly to the full data likelihood, the above posterior contains a high number of product terms. For every item, there is a product over all pairs of individuals. That term is then multiplied by $n$ additional product terms, each of which itself is a product over all pairs of items; However, we may drop all terms that do not contain $\theta_{new}$ and $\mathbf{z}_{new}$ as they are assumed fixed in this sample. Therefore, we now have

\begin{equation}
  \label{eq:personpositions}
\begin{aligned}
  \pi\bigl(\mathbf{z}_{new} | &\mathbf{Y}, \mathbf{U}, \boldsymbol{\beta}, \boldsymbol{\theta},\hat{\boldsymbol{Z}}\bigr) \propto \\ & \pi\left(\mathbf{z}_{new} | \hat{\sigma}_{z}^{2}\right)\prod_{i=1}^{p} \prod_{l = 1}^n \frac{\exp \left(\hat{\beta_{i}}-\left\|\mathbf{z}_{new}-\hat{\mathbf{z}}_{l}\right\|\right)^{y_{i, k t}}}{1+\exp \left(\hat{\beta_{i}}-\left\|\mathbf{z}_{new}-\hat{\mathbf{z}}_{l}\right\|\right)} \prod_{i \neq j} \frac{\exp \left(\theta_{new}-\left\|\hat{\mathbf{w}}_{i}-\hat{\mathbf{w}}_{j}\right\| \right)^{u_{k, i j}}}{1+\exp \left(\theta_{new}-\left\|\hat{\mathbf{w}}_{i}-\hat{\mathbf{w}}_{j}\right\|\right)}
\end{aligned}
\end{equation}

With these conditional posteriors, constructing a Markov Chain Monte Carlo (MCMC) sampler is straightforward and can proceed by providing a random starting value for $\mathbf{z}_{new}$ from the prior distribution and setting a starting value for $\theta_{new}$ by taking the average of $\hat{\theta}_k$ that have the same sum score from the previous sample; for example using Equation \ref{eq:approxtheta} from above. Note that the other when estimating positions for multiple new persons, the new persons do not contribute to each other's estimation as the latent space is assumed fixed and including all new individuals would update the latent space. This is in line with ability estimation practices in IRT. With the exception of online calibration procedures \citep[e.g.,][]{ban2001comparative}, ability and item parameters are not simultaneously updated.

\subsection{Incorporating new items}

This section describes incorporating data from new items into NIRM. These processes all involve fixing some portion of the previously estimated model. In that way, they are similar to various types of fixed-parameter calibration procedures from the IRT literature \citep[e.g,][]{kim2006comparative}. There are two scenarios for incorporating new items. In case 1, some or all of the present sample has been administered new items. This may happen in longitudinal studies or learning environments. In case 2, a new sample has been administered both the complete set of items from the original model plus a new set of items for which new positions must be estimated. Clearly, the former is a more simple endeavor than the latter. In either case, we must update the item intercept using the following conditional posterior.

$$
\pi\left(\beta_{i} | \mathbf{Y}_{i}, \mathbf{Z}\right) \propto \pi\left(\beta_{i}\right) \prod_{k \neq l} \frac{\exp \left(\beta_{i}-\left\|\hat{\mathbf{z}} _{k}-\hat{\mathbf{z}}_{l}\right\|\right)^{y_{i, k l}}}{1+\exp \left(\beta_{i}-\left\|\hat{\mathbf{z}}_{k}-\hat{\mathbf{z}}_{l}\right\|\right)}
$$

Starting values can be obtained in the analogous process to obtaining the starting values for the person intercepts from above. If only item data has been added for the original sample, the person positions used above may be treated as fixed. If the newly added data represents a sample of new persons with response to the old, common items and new ``pre-test'' items, person parameters must be updated using the methods described in the previous section. This process includes a decision point; should the new items contribute to estimating person positions? If yes, then one ends up in a position where persons in the new and old samples with identical response patterns on the common items do not have identical positions in the latent space. If no, the person positions will be identical but the new item positions will only reflect distances from the latent space defined by the original dataset. The former option generally seems more tenable - it makes sense to partially update person positions in light of new information.

Estimating the locations of new items again depends on the type of new data. For case 1, the positions of individuals have been fixed prior to estimation. In case 2, the positions of individuals are estimated concurrently. This somewhat complicates the MCMC sampling scheme, but the procedure for placing the new items positions in the latent space in either case can make use of the conditional posterior in Equation \ref{eq:condpostw} below.

\begin{equation}
  \label{eq:condpostw}
\begin{aligned}
\pi\left(\mathbf{w}_{new} | \mathbf{U}, \boldsymbol{\theta}, \sigma_{w}^{2}\right) \propto \pi\left(\mathbf{w}_{new} | \hat{\sigma}_{z}^{2}\right) & \prod_{i=1}^{p} \prod_{k \neq l} \frac{\exp \left(\beta_{i}-\left\|\mathbf{z}_{k}-\mathbf{z}_{l}\right\|\right)^{y_{i, k t}}}{1+\exp \left(\beta_{i}-\left\|\mathbf{z}_{k}-\mathbf{z}_{l}\right\|\right)} \\ & \prod_{k=1}^{n} \prod_{i \neq j} \frac{\exp \left(\hat{\theta}_{k}-\left\|\hat{\mathbf{w}}_{i}-\hat{\mathbf{w}}_{j}\right\| \right)^{u_{k, i j}}}{1+\exp \left(\hat{\theta}_{k}-\left\|\hat{\mathbf{w}}_{i}-\hat{\mathbf{w}}_{j}\right\|\right)}
\end{aligned}
\end{equation}

A similar approach to that in Equation \ref{eq:personpositions} can be applied to limit the number of calculations in the sampling procedure. Dropping all pairs of items that do not contain at least one new item results in placing the new items in the latent space. It is possible to further simplify computation by only including pairs of items where exactly one item is new. The distinction between the former and the latter equates to not updating the latent space at all vs. performing a partial update of the item latent space in which the positions of the new items take both new and old items into account. Care must be taken when deciding between these two options. If the item latent space is not updated and new items are simply ``placed'' (or ``positioned'') into the old latent space, then 1.) the distance between the locations of the new items may not be an accurate reflection of their true distances and 2.) clusters (or lack-thereof) of new items may be difficult to evaluate. Generally, it should be preferred to partially update the latent space with the new items.

\section{Illustrative Data Analysis}

In this section, we illustrate the workflow of estimating a NIRM of student assessment data. The R code used in the application are provided in the online repository associated with this study.

\subsection{Data Background}
The AP-CAT (Advanced Placement - Computerized Adaptive Testing) project was funded by the National Science Foundation to develop formative assessments for high-school AP statistics classes. This project required item bank calibration, development of an online system for data collection, pilot testing of multiple fixed-form and CAT assessments with the ultimate goal of
providing granular diagnostic feedback to students via detailed score reports. During the pilot phases of the project, students in several classrooms took up to 5 assessments throughout spaced within year-long AP statistics course. The item development phase included an expert panel who associated items with the content areas of AP statistics, which at that point in time had 4 content domains.

This study used data from the 2018-19 cohort of the AP-CAT Project. In total, that sample has a total size of 441 participants. All participants completed appropriate parental consent and assent forms to obtain eligibility for participation in this study. Participants received access to the AP-CAT platform at no cost as part of this study. As part of the study, students completed several self-report surveys on learning-related constructs and a set of linear test assignments that contributed to course grade. The subsample (N = 368) of those who completed the standardized assignment at the second time point were included in the analysis.

\subsection{Measure}

The assessment used in this illustrative example contained 27 common items across all teachers whose classrooms participated in this study; some teachers also taught multiple sections. Some teachers elected not to administer all items due to the classrooms having not covered the related content at that time point. The assessment contained items from 3 of the 4 main topic areas, the majority balanced between area 2 and 3 as content area 1 had been assessed in the assessment administered at the 1st time point. The breakdown of item content area is further described in Table \ref{tab:content}, and example items are provided in the online respository.

\tablehere{\ref{tab:content}}

\subsection{Analysis}

To analyze the data, we assume the distances between items and persons can be approximated with a latent space comprising 2 dimensions; this was selected both for ease of visualization and due to the majority of the items coming from two content domains. It is not necessary for the dimension of the latent space dimension to match any ``true'' number of dimensions. Rather, increasing the number of dimensions would ease post-hoc detection of item clusters (e.g. by spectral clustering of the latent space) by increasing the number of directions in which separation may occur. This is a tradeoff that comes at the cost of limiting visualization of the clusters in the latent space.

We selected the ``all concordant pairs'' specification (Equation \ref{eq:pow}) of the item networks rather than the ``positive concordant pairs'' (\ref{eq:prod}) representation because it was believed that absence of a correct response indicates a lack of proficiency in the attributes required to respond to the question. Thus, it is desirable if incorrect response would contribute to an increased distance from that item's position. The linkage between the networks was chosen such that the person space is a function of the item latent space (or equivalently, the item positions). This selection was informed by the adequately large number of items, allowing for person positions to reflect a large degree of variability in the individual patterns of response. The Bayesian estimation procedure outlined in \citet[][pp. 244]{jin2019doubly} and the Estimation section above. The parameters of the prior distribution were chosen that the intercept parameters had a prior mean of 0 and a prior variance of $\sigma_\theta$ = $\sigma_\beta$ = 10. We elected to estimate the contributions of the latent spaces by placing an Inv-Gamma hyperprior on the variability of the latent spaces, with $a_\sigma = b_\sigma = .001$

\subsection{Results}

The analysis took 39 minutes to complete on a using 4 threads of a 4 core Intel i7-4810 CPU @ 2.80GHz with 16GB of RAM to run a change of length 15000 of with the first 5000 were discarded for burn-in and a thinning rate of 5 for 2000 total samples. Posteriors distribution summaries for person intercept parameters are groups by sum score and are provided in Table \ref{tab:theta}. 

\tablehere{\ref{tab:theta}}

$\theta$ estimates have a curvilinear relationship with the sum score because when the ``all concordant pairs'' network specification is used, $\theta$ represents consistency of the pattern rather than the pattern itself. Similarly, $\beta$ estimates are provided in \ref{tab:beta}, and tend to increase as the proportion correct on the items moves away from the center.

\tablehere{ \ref{tab:beta}}

\begin{figure}[H]
\centerline{\includegraphics[scale=.3]{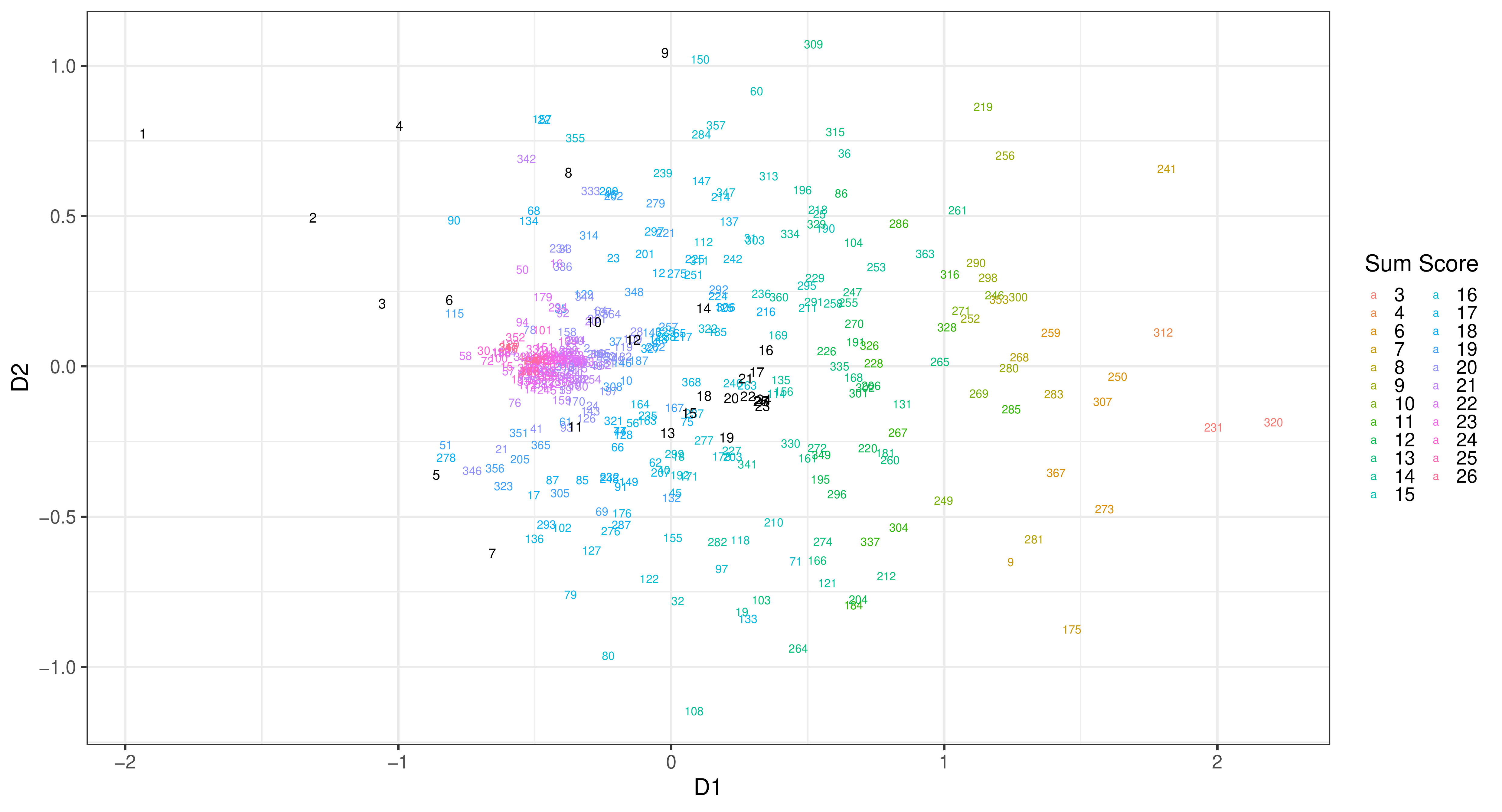}}
    \caption{Item and person positions for the illustrative data example. Items are labeled in black, individuals are labeled with an arbitrary ID number and color coded by sum score.}
  \label{fig:personpos}
 \end{figure}

Person positions are displayed in Figure \ref{fig:personpos}, with item positions superimposed.

\figurehere{\ref{fig:personpos}}

The first thing to notice about the person positions plot is the single large cluster of person positions near the center of the figure. There participants all have highly similar sum scores (represented by color) and form a group with very small distances to each other. For example, participants number 30 and number 352 are on the upper left edge of the cluster in the center. Because of the proximity, they likely have very similar response patterns. However, given that they have the same sum score (24, denoted by the same color), but different positions, these participants must have slight differences in their response patterns. Indeed, the pattern for participant 30 is ``111110101111111011111111111'' and the pattern for participant 352 is ``011111111101111111011111111''.  The differ on 6 items, $\{1,6, 8, 11, 16, 19\}$; the pattern score for only those 6 items is "011010" for participant 30 and  "100101" for participant 352.

\tablehere{\ref{table:seveneightnine}}
The next thing to inspect is items the contrast between items that have large distances from one another. For example, items 7 and 9 are on the opposite side of the plot about the x axis. Item 7 had 193 participants answer the item correctly and item 9 had 195 participants answer the item correctly. The proportions correct are very similar, but the two items were not estimated to reside in similar locations. This can be easily understood from viewing the $2 \times 2$ contingency table for the items. The table shows there are 170 concordant pairs and 198 discordant pairs. Comparing this to the number of concordant pairs of item 8 and 9, which has 193 concordant pairs while item 8 has 194 correct responses. This three way simplification demonstrates how proximity reflects patterns in the $2 \times 2 \times \ldots 2$ contingency tables that may be difficult to spot without a high degree of effort. Table \ref{table:seveneightnine} displays the three-way contingency table for items 7, 8 and 9. 

\begin{figure}[H]
\centerline{\includegraphics[scale=.35]{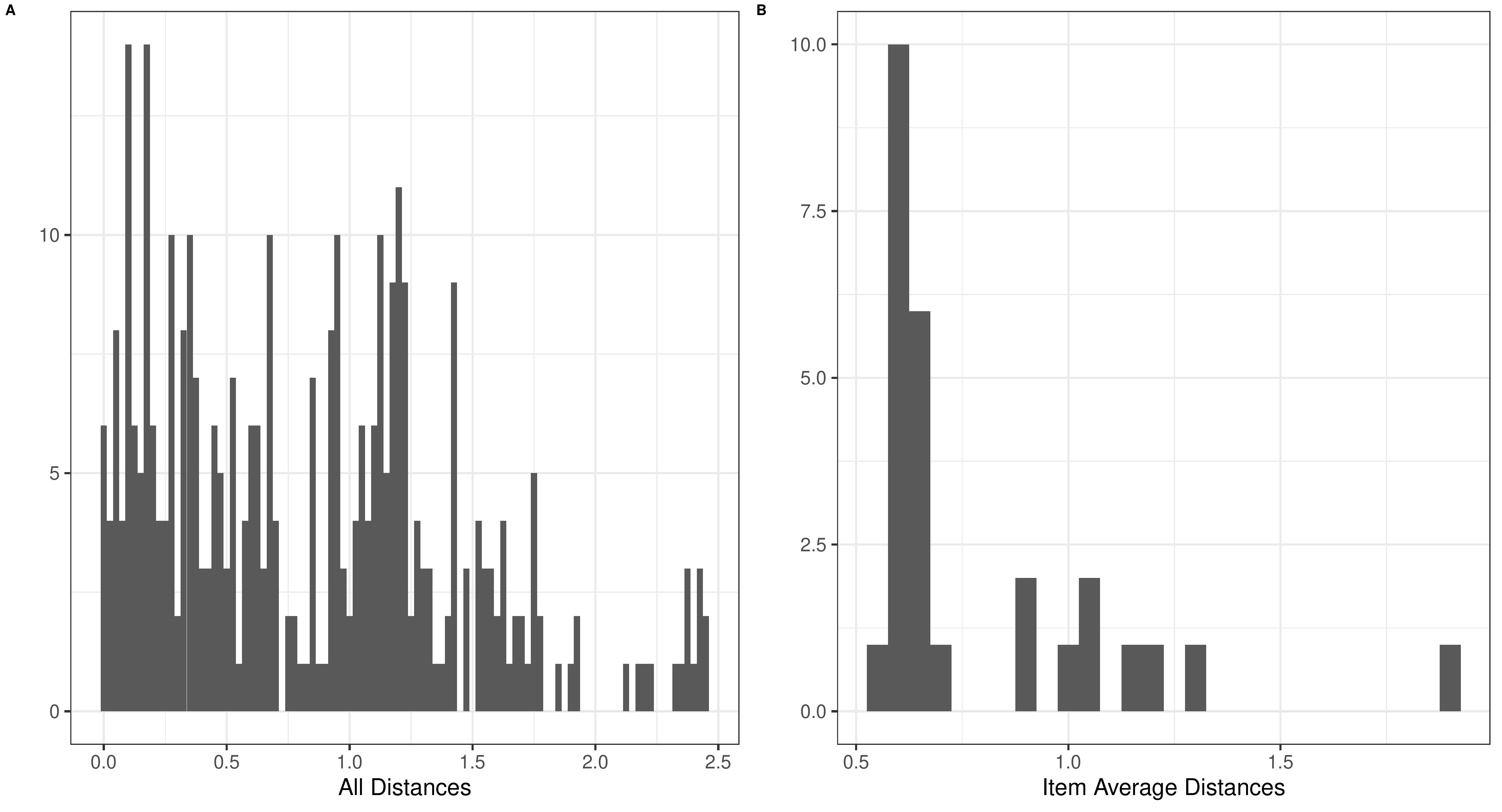}}
  \caption{A: Histogram of all $\binom{p}{2}$ distances in the latent space. The multiple peaks around .4 and 1.2 suggest there are multiple item clusters with different degrees of spread. B: Histogram of the average distance between an item and the other items. Item 1 has the largest average distance at 1.9 }
  \label{fig:bothhist}
\end{figure}

\figurehere{\ref{fig:bothhist}}

Next, we examine for outlier items in the latent space. From visual inspection, it appears that item 1 has the largest distances from the other items. However, this is easily verified by calculating the average distances from every item to every other item. For example, in R code this would be \texttt{colMeans(dist(W))} where \texttt{W} is the matrix containing the item positions. Indeed, item 1 has the largest average distance to the other items at 1.9 units. Pane B of Figure \ref{fig:bothhist} contains a histogram of average item-rest distances. Pane A of the same figure displays a histogram of all $\binom{p}{2}$ distances. The large distance of item 1 to the rest of the items is cause for suspicion into the items fit within the rest of the items. The item with the largest distance from item 1 is item 23; those two have 281 discordant pairs and only 86 concordant pairs. Here we may attempted to infer the source of these differences. Given an educational test, we suspect that two possible causes of the low connectedness between those two nodes may be due to different content areas or potentially a miskeyed response. Inspecting the items, both items 1 and 23 are from content area 3. Thus it is not likely that the items have different patterns do to different knowledge requirements. However, the second possibility was found to be plausible - item 1 contains multiple response options which may be considered to be correct by the test takers.

\begin{figure}[H]
\centerline{\includegraphics[scale=.5]{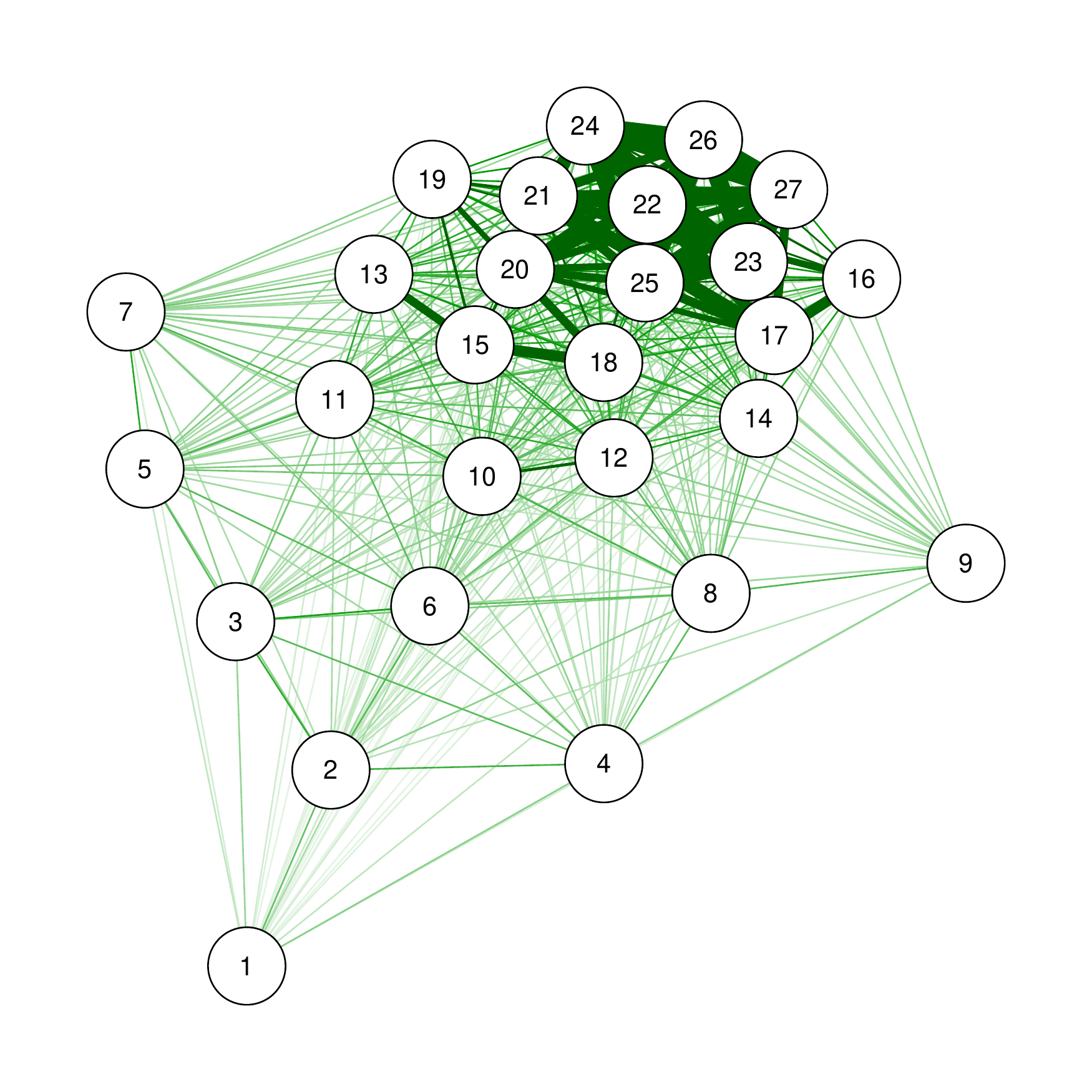}}
    \caption{Item and person positions for the illustrative data example. Items are labeled in black, individuals are labeled with an arbitrary ID number and color coded by sum score.}
  \label{fig:illnetwork}
 \end{figure}

\figurehere{\ref{fig:illnetwork}}

Figure \ref{fig:illnetwork} displays the network of item similarities (determined by Inverse-Exponentiated distances, $s_1$). In this figure, positions in the network are determined by choosing \texttt{layout = "spring"} option in the \texttt{qgraph} function in the R package of the same name \citep{epskamp2012qgraph}. The highly similar items are all also highly interconnected, with many of items that have at least one high value of similarity tending to have many additional high similarity connections. The next section will compare NIRM to two other commonly used psychometric models, and fit versions of these models to the same dataset used in this illustrative example.

\section{Comparisons to Other Models}

To compare and contrast Item Response Theory and the Ising Model with the network item response model, we first briefly describe both.

\subsection{Item Response Theory}

We first consider the multidimensional version of the 2 parameter logistic IRT model. In this model the probability of a student yielding a correct response is:
$$P(X_{ki}=1|\mathbf{\theta}) = \frac{1}{1 + exp(-(\mathbf{\alpha}^\intercal \mathbf{\theta} + \delta))}$$

Here, $\mathbf{\theta}$ represents the latent ability vector, $\mathbf{\alpha}$ is the vector of discrimination parameters, and $\delta$ is the item intercept. For typical models, the full data likelihood for $n$ persons and $p$ items has the following form:

$$\prod_{k=1}^{n} \prod_{i=1}^{p} (\frac{1}{1 + exp(-(\mathbf{\alpha}_{i}^\intercal \mathbf{\theta}_{k} + \delta_{i}))})^{x_{ki}}(1-\frac{1}{1 + exp(-(\mathbf{\alpha}_{i}^\intercal \mathbf{\theta}_{k} + \delta_{j}))})^{1-x_{ki}}$$

Within the form of the likelihood, i.e. the double product over items and persons, we find the baked in assumptions of independent persons (e.g. random sampling) and conditional independence of items. The latter of these two is the well known local independence assumption. This set of assumptions is quite strict in that it requires explicitly adding terms to the model in order to account for various types of dependency. See, for example, mixture IRT for unobserved person clusters \citep{rost1990rasch}, multiple group IRT for known person clusters \citep{bock1997multiple}, bifactor models for known item clusters caused by a known construct \citep{gibbons2009psychometric}, testlet models for known item clusters caused by an unknown construct \citep{demars2006application}, and etc. Clearly, there is no shortage of techniques to accommodate independence assumption violations in IRT. 

\subsection{Latent Ising Model}

In the binary networks used in the Ising model, variables $x_i$ take values $\in {-1,1}$ as opposed to $\in {0,1}$ due to the traditional use in physics where the network nodes represented spins on a lattice. The Ising model as used in psychological research, has the following form:
$$\begin{aligned}
        p(\mathbf{x}) &=p\left(x_{1}, x_{2}, \ldots, x_{n}\right) \\
        &=\frac{\exp \left\{\sum_{i=1}^{n} \mu_{i} x_{i}+\sum_{<i, j>} \sigma_{i j} x_{i} x_{j}\right\}}{\sum_{\mathbf{x}} \exp \left\{\sum_{i=1}^{n} \mu_{i} x_{i}+\sum_{<i, j>} \sigma_{i j} x_{i} x_{j}\right\}}
        \end{aligned}
        $$

where the summation over$<i,j>$ is over all pairs of nodes $(i,j)$ that are connected neighbors on  graph, and $\sum_{\mathbf{x}}$ is the sum over all possible $2^n$ patterns of the binary variables. Nodes are considered to be independent conditional on the values of the other nodes which they are ``connected''. In traditional Ising models, only nodes that were adjacent on the lattice were considered to be connected. However, in psychometric networks it is desirable to consider all nodes to be eligible for connection, and then estimate which nodes are connected and to what degree. The model above contains main effects and interaction terms. The main effects determine the tendency for node $x_i$ to take positive (if $\mu_i > 0$) or negative values (if $\mu_i < 0$). If two nodes have a positive interaction ($\sigma_{ij} > 0$) then those nodes will have a tendency to take the same values. Likewise, negative interaction terms ($\sigma_{ij} < 0$) represent a tendency for items to take opposite values. Importantly, these terms are interpreted similarly to regression coefficients in that the effect only has meaning at fixed values of other variables. Stated differently, the interpretation of the interaction term for $x_i$ and $x_j$ has the preceding interpretation conditional on the values of the other nodes.

\subsection{Comparison}

Without discussing further here, we note that \citet{marsman2018introduction} provided a clever linkage between Ising models and multi-dimensional 2 parameter logistic models that shows each Ising model has a statistically equivalent IRT model. While true, \citet{epskamp2017estimating} make clear that interpretations and implications of each model are drastically different. IRT models represent a \textit{common cause} model, in which the responses to items are caused by common underlying factor(s). Conversely, Ising models represent another paradigm in which observable features form a network of nodes connected by causal relations. However, both models are effectively multivariate generalized linear models. In IRT, it is latent variables that are used to predict the outcome. In the Ising model, all variables and all possible interactions of variables are used as predictors. 

NIRM, on the other hand, does not attempt to predict the response value of a single item and instead it chooses some collection of elements of the $2 \times 2$ contingency tables for all items/persons to predict. From this perspective, NIRM is attempting to answer a different question than either IRT or Ising models: What is an optimal representation of distances between item and person positions, and conditional on marginal tendencies, that best replicates the observed summaries of pairwise responses. In particular, for the second of the two pairwise encoding procedures discussed, there is no way to recover the data in row-column form once the networks have been constructed; that is concordant pairs of all types are treated equally and all discordant pairs are treated equally. This understanding of how NIRM, the Ising model, and IRT conceptualize pairwise responses is key to understanding their differences and is discussed more deeply in the next section. 

\subsection{Response Pattern Similarities}

One driver in the development of network approaches in psychometrics is the pursuit of alternative explanations of positive correlations among tests of ability. That is, the prevailing models that account for the so called positive manifold \citep{spearman1925some} of responses is contain a scalar/vector of latent trait scores that cause all response patterns. On the other hand, the Ising model explains the positive manifold as a consequence of ``a network of mutually reinforcing entities connected by causal relations'' \citep{marsman2018introduction}. Both Ising models or IRT models imply an the explanation of why response patterns may be similar to one another. These explanations may also be very similar. If binary response patterns are the same for two people, the IRT explanation would be that it is likely the two individuals have similar values for their latent trait scores, and given that, they independently generated similar response patterns. Similarly, for two identical patterns generated from an Ising model perspective, the explanation would be that due to the connections interactions present in the network, some patterns are more likely than others. Then, identical patterns are a product of independent draws from a categorical distribution with non-equal probabilities.

Considering another angle, why might two items have a large number of concordant pairs? In IRT, the number of concordant pairs from two items is solely function of the item parameters. That is for a fixed sample size N, the number of concordant pairs can be written as

\begin{align*}
  N*p_{11} + N*p_{00} = \\& N * (P(U_1 = 1 \cap U_2 = 1) + P(U_1 = 0 \cap U_2 = 0)) =\\ &N * \big( \int_{-\infty}^{\infty} P(U_1 = 1|\theta)P(U_2 = 1|\theta)f(\theta) d\theta + \\ &  \int_{-\infty}^{\infty} (1-P(U_1 = 1|\theta))(1-P(U_2 = 1|\theta))f(\theta) d\theta \big)
\end{align*}

In the Ising model, the number of concordant pairs for a set of items is a function of $\mu_i x_i + \mu_j x_j + \sigma_{ij} x_i x_j$ and implies that this probability is the same for all possible patterns of the other variables. This implication is a byproduct of the absence of latent variables in the model, or rather, the Ising model representing a pairwise markov field of the binary items. In NIRM, similar response patterns are caused by proximity of persons in the Euclidean latent space. Technically, the distance between each person contributes to the pairs of responses. All said, the Ising model and IRT models result in different findings when fit to a dataset. To make the distinction clear, we will estimate the Ising Model and a unidimensional 2PL model on the same data from the illustrative example. The Ising model was estimated with the \texttt{IsingFit} function (using default settings) in the package of the same name \citep{bork2016ising}. 

\begin{figure}[H]
\centerline{\includegraphics[scale=.5]{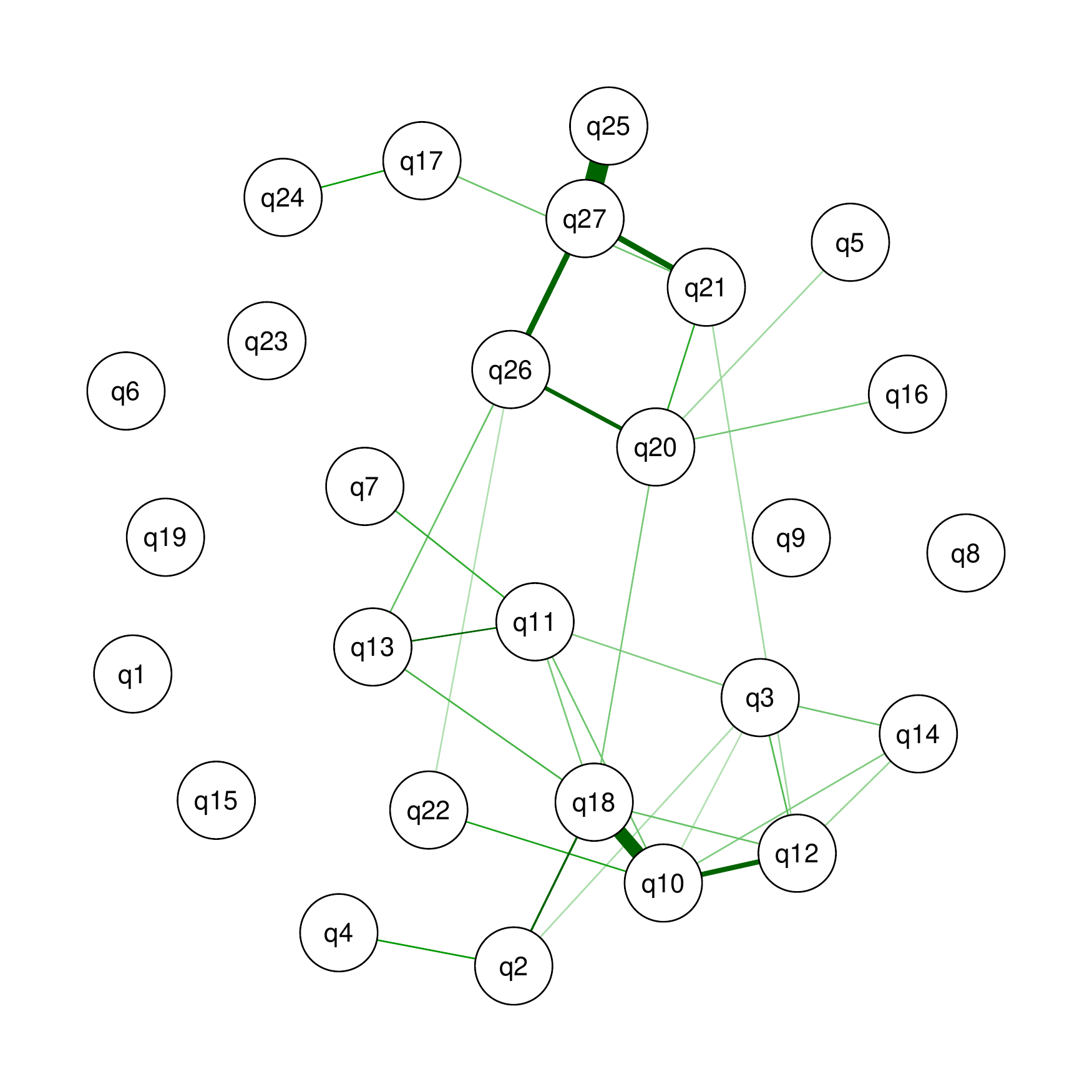}}
  \caption{Network visualization of the Ising Model.}
  \label{fig:isingassess}
\end{figure}

As previously mentioned, Ising models are frequently visualized via a networked plot of the nodes, which is displayed in Figure \ref{fig:isingassess}. This estimated model has many unconnected nodes, with only a few clusters of connections. The estimated thresholds for the IsingModel are in Table \ref{tab:comparisonparms}. The interpretation of the Ising Model visualization is quite different than the NIRM model. In the Ising model, the connections represent a tendency to covary after taking into a account the \emph{values} of all other variables. In contrast, in NIRM, a connection represents a tendency to have a relatively large number of concordant pairs after taking into account the \emph{positions} of all other variables. 

\figurehere{\ref{fig:isingassess}}

The 2PL model was estimated using the \texttt{mirt} package in R \citep{chalmers2012mirt}. The loadings and intercepts are also provided in Table \ref{tab:comparisonparms}. An interesting note that there are some red flags in the values of the parameter estimates for the IRT model. First, Item 1 has a negative (yet rather close to zero) factor loading. For educational data, this is rarely expected; any in-depth analysis would probe into the data further to understand this issue. Similarly, Item 27 has an unusually large intercept (6.83), reflective of the near-perfect proportion correct of the item (.98). For large-scale testing programs, this item would be considered far to easy to provide useful information in discarded. Both of the findings from the IRT model were corroborated by evidence from the NIRM. However, the NIRM and the Ising Model led to somewhat different substantive conclusions. This is not to say that either conclusion is wrong, but the information received is certainly different between the two. 

\tablehere{\ref{tab:comparisonparms}}

\section{Discussion and Conclusions}

This study intended to show application of Network Item Response Models to student assessment data by introducing the model and providing in-depth discussion of the multiple decision points that occur throughout the model building procedure. However, this discussion is not complete as some questions surrounding network item response models are currently unanswered. One open question in NIRM is a procedure for evaluating model fit. Given the Bayesian estimation procedure, posterior predictive checks are always possible. However, more standardized metrics would be desirable such that cut-off values can be determined. While the quantification of the latent spaces via estimating $\sigma^2_z$ can capture the contribution, it is unknown how this metric response to increases in dimension, number of items, and number of persons. A similar question to model fit is the sample size requirements that are necessary for estimating NIRM. The present study had a sample size of slightly less than 400. The first study on NIRM \citep{jin2019doubly} analyzed a real dataset that had roughly 300 participants and provided a simulation study comparing NIRM to a mixture IRT model. In that study, at least, a sample of 300 was adequate enough to outperform mixture IRT in detecting person clusters. However, more research is needed.

There are several other open questions in NIRM. For example, this manuscript provided two specifications for linkage between the latent spaces and two specifications for constructing the items networks. The differences between the 4 combinations of these choices have not been evaluated via simulation study. Two further concepts that may require further study are identification and regularization. It has been stated multiple times that the positions in NIRM are inherently meaningless. However, it may be possible to identify the locations in the model by fixing the positions of a small number of response patterns (e.g. d + 1) to certain positions in the latent space. For example, identifying the response pattern of all 1's to occur at the point $\mathbf{0}^d$. Furthermore, estimators of Euclidean distance may be positively biased when the true distance is in the neighborhood of 0. This may be concerning when attempted to cluster items, as inflated estimates of distance make the clustering processes more difficult. A possible solution is to use a prior of the form described by \citet{madigan1994model}, and this may also be a direction of future study. 

We have shown that modeling networks constructed from item responses in student assessment data can provide interesting insights into the structure of the data. First, it is a useful procedure for detecting items that do not display strong conformance to other items. The modelling procedure is particularly useful due to the ease with which the resulting latent spaces can be visualized. Comparing to an IRT model in which one must calculate summary information about parameter estimates, clusters of individuals and individual outliers are easily spotted from visual inspection of the latent space plots. Interestingly, even though under specific conditions IRT models are equivalent to Ising models, in this case they let to different conclusions. 

\vspace{\fill}\pagebreak



\bibliographystyle{apacite}
\bibliography{references}


\vspace{\fill}\pagebreak

\section*{Figures}

\vspace{\fill}\pagebreak

\section*{Tables}

\begin{table}[H]
  \centering
  \begin{tabular}{|c|c|c|c|}
    \hline 
    & $X_{li}=0$ & $X_{li}=1$\\ 
    \hline 
    $X_{ki}=0$ & $a$ & $b$\\
    \hline 
    $X_{ki}=1$ & $c$ & $d$\\
    \hline
  \end{tabular}
  \caption{$2 \times 2$ contingency table. Positively matched concordant pairs are represented by $a$, all concordant pairs are $a + b$, etc. }
  \label{tab:otu}
\end{table}

  \begin{table}[H]
  \centering
  \begin{tabular}{rr}
  \hline
  qid & Content Area \\ 
  \hline
  6 & 1 \\ 
  8 & 1 \\ 
  2 & 2 \\ 
  3 & 2 \\ 
  4 & 2 \\ 
  10 & 2 \\ 
  11 & 2 \\ 
  12 & 2 \\ 
  13 & 2 \\ 
  14 & 2 \\ 
  17 & 2 \\ 
  19 & 2 \\ 
  20 & 2 \\ 
  23 & 2 \\ 
  24 & 2 \\ 
  26 & 2 \\ 
  1 & 3 \\ 
  5 & 3 \\ 
  7 & 3 \\ 
  9 & 3 \\ 
  15 & 3 \\ 
  16 & 3 \\ 
  18 & 3 \\ 
  21 & 3 \\ 
  22 & 3 \\ 
  25 & 3 \\ 
  27 & 3 \\ 
  \hline
  \end{tabular}
\caption{Item Content Areas and question id (qid).}
\label{tab:content}
\end{table}

\begin{table}[H]
\centering
\begin{tabular}{cc*{2}{S[table-format=2]}} 
\toprule 
& & \multicolumn{2}{c}{Item 9} \\ 
\cmidrule{3-4} 
\multirow{-2.5}{*}{\makecell{Item \\ 7}} & \multirow{-2.5}{*}{\makecell{Item \\ 8}} & {0} & {1}\\ 
\midrule 
0 & 0 & 43& 50\\ 
&1 & 32&    50\\ \addlinespace
1&0 & 43&    38\\ 
& 1 & 55&  57\\ 
\bottomrule 
\end{tabular}
\caption{2 way contingency table for items 7, 8, and 9}
\label{table:seveneightnine}
\end{table}

\begin{table}[H]
\centering
\begin{tabular}{rrrrrrr}
  \hline
sum\_score & freq & min & 25\% & med & 75\% & max \\ 
  \hline
3.00 &   2 & 1.87 & 2.19 & 2.28 & 2.38 & 2.76 \\ 
  4.00 &   1 & 1.61 & 1.85 & 1.95 & 2.03 & 2.32 \\ 
  6.00 &   5 & 1.08 & 1.38 & 1.46 & 1.53 & 1.81 \\ 
  7.00 &   3 & 0.93 & 1.20 & 1.27 & 1.34 & 1.63 \\ 
  8.00 &   4 & 0.74 & 1.04 & 1.11 & 1.19 & 1.47 \\ 
  9.00 &   6 & 0.63 & 0.91 & 0.99 & 1.07 & 1.37 \\ 
  10.00 &   5 & 0.53 & 0.81 & 0.90 & 0.97 & 1.27 \\ 
  11.00 &  10 & 0.45 & 0.74 & 0.81 & 0.89 & 1.17 \\ 
  12.00 &  12 & 0.43 & 0.69 & 0.77 & 0.84 & 1.12 \\ 
  13.00 &  20 & 0.37 & 0.66 & 0.74 & 0.81 & 1.10 \\ 
  14.00 &  19 & 0.39 & 0.67 & 0.74 & 0.82 & 1.10 \\ 
  15.00 &  13 & 0.38 & 0.69 & 0.77 & 0.84 & 1.13 \\ 
  16.00 &  25 & 0.46 & 0.74 & 0.82 & 0.89 & 1.18 \\ 
  17.00 &  24 & 0.52 & 0.81 & 0.89 & 0.96 & 1.25 \\ 
  18.00 &  38 & 0.63 & 0.91 & 0.99 & 1.07 & 1.35 \\ 
  19.00 &  28 & 0.76 & 1.04 & 1.11 & 1.19 & 1.50 \\ 
  20.00 &  34 & 0.89 & 1.19 & 1.27 & 1.34 & 1.63 \\ 
  21.00 &  26 & 1.08 & 1.37 & 1.45 & 1.53 & 1.83 \\ 
  22.00 &  36 & 1.31 & 1.59 & 1.67 & 1.75 & 2.06 \\ 
  23.00 &  26 & 1.55 & 1.86 & 1.95 & 2.03 & 2.37 \\ 
  24.00 &  13 & 1.86 & 2.19 & 2.28 & 2.38 & 2.75 \\ 
  25.00 &  12 & 2.28 & 2.65 & 2.75 & 2.86 & 3.29 \\ 
  26.00 &   6 & 2.88 & 3.39 & 3.53 & 3.67 & 4.24 \\ 
   \hline
\end{tabular}
\caption{Theta Estimates}
\label{tab:theta}
\end{table}

\begin{table}[H]
\centering
\begin{tabular}{rrrrr}
  \hline
item & $\hat{p}$ & $\hat{\beta}$ & 5\% & 95\% \\ 
  \hline
  1 & 0.09 & 2.54 & 2.52 & 2.56 \\ 
    2 & 0.29 & 1.14 & 1.12 & 1.15 \\ 
    3 & 0.38 & 0.89 & 0.88 & 0.91 \\ 
    4 & 0.40 & 0.85 & 0.83 & 0.86 \\ 
    5 & 0.50 & 0.77 & 0.75 & 0.79 \\ 
    6 & 0.50 & 0.77 & 0.75 & 0.79 \\ 
    7 & 0.52 & 0.78 & 0.76 & 0.79 \\ 
    8 & 0.53 & 0.78 & 0.76 & 0.79 \\ 
    9 & 0.53 & 0.78 & 0.76 & 0.79 \\ 
   10 & 0.60 & 0.86 & 0.84 & 0.87 \\ 
   11 & 0.61 & 0.88 & 0.86 & 0.90 \\ 
   12 & 0.66 & 1.00 & 0.98 & 1.02 \\ 
   13 & 0.71 & 1.17 & 1.15 & 1.19 \\ 
   14 & 0.72 & 1.20 & 1.18 & 1.22 \\ 
   15 & 0.74 & 1.26 & 1.24 & 1.27 \\ 
   16 & 0.77 & 1.42 & 1.40 & 1.44 \\ 
   17 & 0.78 & 1.46 & 1.44 & 1.48 \\ 
   18 & 0.78 & 1.48 & 1.46 & 1.50 \\ 
   19 & 0.79 & 1.49 & 1.47 & 1.51 \\ 
   20 & 0.79 & 1.51 & 1.49 & 1.53 \\ 
   21 & 0.79 & 1.54 & 1.52 & 1.56 \\ 
   22 & 0.83 & 1.74 & 1.72 & 1.76 \\ 
   23 & 0.84 & 1.81 & 1.80 & 1.83 \\ 
   24 & 0.88 & 2.15 & 2.13 & 2.17 \\ 
   25 & 0.91 & 2.54 & 2.52 & 2.56 \\ 
   26 & 0.94 & 2.96 & 2.93 & 2.98 \\ 
   27 & 0.98 & 4.17 & 4.13 & 4.21 \\ 
   \hline
\end{tabular}
\caption{Beta Estimates}
\label{tab:beta}
\end{table}

\begin{table}[H]
\centering
\begin{tabular}{r|rr|r}
  \hline
  \hline
  &\multicolumn{2}{|r|}{IRT Model}& Ising Model\\
 Item & $\alpha$ & $\delta$ & $\mu$ \\ 
  \hline
1 & -0.07 & -2.35 & -2.35 \\ 
  2 & 1.18 & -1.11 & -2.24 \\ 
  3 & 1.64 & -0.72 & -1.44 \\ 
  4 & 0.44 & -0.40 & -0.57 \\ 
  5 & 0.59 & 0.01 & -0.22 \\ 
  6 & 0.78 & 0.01 & 0.01 \\ 
  7 & 0.56 & 0.10 & -0.31 \\ 
  8 & 0.80 & 0.12 & 0.11 \\ 
  9 & 0.08 & 0.12 & 0.12 \\ 
  10 & 1.95 & 0.65 & -2.25 \\ 
  11 & 1.52 & 0.65 & -1.66 \\ 
  12 & 1.84 & 1.04 & -0.40 \\ 
  13 & 1.22 & 1.17 & 0.16 \\ 
  14 & 1.10 & 1.18 & 0.38 \\ 
  15 & 1.28 & 1.33 & 1.03 \\ 
  16 & 0.75 & 1.36 & 0.34 \\ 
  17 & 0.88 & 1.46 & 0.13 \\ 
  18 & 1.92 & 2.00 & -0.92 \\ 
  19 & 0.99 & 1.54 & 1.30 \\ 
  20 & 1.55 & 1.84 & -0.27 \\ 
  21 & 1.22 & 1.71 & -0.20 \\ 
  22 & 1.39 & 2.07 & 0.92 \\ 
  23 & 0.80 & 1.83 & 1.64 \\ 
  24 & 0.71 & 2.15 & 1.57 \\ 
  25 & 1.20 & 2.88 & 0.18 \\ 
  26 & 1.80 & 3.96 & 0.67 \\ 
  27 & 2.66 & 6.83 & 0.44 \\ 
   \hline
\end{tabular}
\caption{(A.) Item Parameter Estimates from a Unidimensional IRT model. The model converged after 20 EM iterations. Note that some parameter values are extremely outside the range of normal variation of item parameters ($\delta \approx \in [-2, 2]$ and $\alpha > 0$).  (B.) Threshold Parameters for the Ising Model}
\label{tab:comparisonparms}
\end{table}


\end{document}